# The Connection Between Inertial Forces and the Vector Potential

Alexandre A. Martins[1] and Mario J. Pinheiro[2]

[1]*Condensed Matter Physics Center, University of Lisbon, Lisboa, Portugal*
[2]*Department of Physics and Center for Plasma Physics, & Instituto Superior Técnico, Lisboa, Portugal*
*351.1.21.841.93.22, mpinheiro@ist.utl.pt*

**Abstract.** The inertia property of matter is discussed in terms of a type of induction law related to the extended charged particle's own vector potential. Our approach is based on the Lagrangian formalism of canonical momentum writing Newton's second law in terms of the vector potential and a development in terms of obtaining retarded potentials, that allow an intuitive physical interpretation of its main terms. This framework provides a clear physical insight on the physics of inertia. It is shown that the electron mass has a complete electromagnetic origin and the covariant equation obtained solves the "4/3 mass paradox". This provides a deeper insight into the significance of the main terms of the equation of motion. In particular a force term is obtained from the approach based on the continuity equation for momentum that represents a drag force the charged particle feels when in motion relatively to its own vector potential field lines. Thus, the time derivative of the particle's vector potential leads to the acceleration inertia reaction force and is equivalent to the Schott term responsible for the source of the radiation field. We also show that the velocity dependent term of the particle's vector potential is connected with the relativistic increase of mass with velocity and generates a stress force that is the source of electric field lines deformation. This understanding broadens the possibility to manipulate inertial mass and potentially suggests some mechanisms for possible applications to electromagnetic propulsion and the development of advanced space propulsion physics.

## INTRODUCTION

There have been numerous attempts to explain the origin of the inertia property of matter suggested qualitatively by Galileo in his writings and later quantified by Newton (Jammer, 1961), although its conceptualization still remains as an unclear resistance of mass to changes of its state of motion. Since the works of Kirchhoff, Mach, Hertz, Clifford and Poincaré a set of logical objections were raised against Newton's laws mainly based on the significance of mass and force (Eisenbud, 1958). Driven by a strong need to develop advanced space propulsion physics there are studies that try to link inertia with gravitational interactions with the rest of the universe (Mach, 1989; Bridgman, 1961; Sciama, 1953), while others hypothesize that inertial forces result from an interaction of matter with electromagnetic fluctuations of the zero point field (Sakharov, 1968; Puthoff, 1989; Haisch and Rueda, 1998) and, finally, others that attribute inertia as the result of the particle interaction with its own field (Lorentz, 1992; Abraham, 1902; Richardson, 1916; see also Ray (2004) for a general survey).

Historically, it was the experimental studies of electrically charged particles animated by high velocities that lead physicists to introduce the notion of electromagnetic inertia besides mechanical inertia. J. J. Thomson was the first to introduce the idea of a supplementary inertia with constant magnitude for a charge $q$ with radius $R$ in a medium of magnetic permeability μ, to be summed up with the mechanical mass $m$ such that $m + 4\mu q/(15R)$ (Thomson, 1881) (see Arzeliés (1966) for a deep historical account). Inspired probably on Stokes (1849) finding that a body moving in water seems to acquire a supplementary mass, Thomson built a hydrodynamical model with tubes of force displacing the ether. However, these studies have not achieved a clear and concise explanation of the phenomenon although different approaches to the classical model of the electron in a vacuum may contribute to clarify hidden aspects of the problem (Dehmlet, 1989). Until now there is no experimental support of Mach's principle as a recent experimental test using nuclear-spin-polarized $9Be^+$ ions gives null result on spatial anisotropy and thus supporting local Lorentz Invariance (Prestage et al., 1985). This supports our viewpoint that inertia is a local phenomena and it is along this epistemological basis that we discuss the inertia property of matter in terms of an interaction of material particles own vector potential with mechanical momentum. After all, quantum electrodynamics was built on a similar basis (Tomonoga (1966) in his Nobel lecture describes the process on the following grounds: "The electron, having a charge, produces an electromagnetic field around itself. In turn, this field, the so-called self-field of the electron, interacts with the electron [...] Because of the field reaction the apparent mass of the electron differs from the original mass"). However, it is clear from this account that not all electromagnetic mass has an electromagnetic nature in QED phenomenology.

Limiting our considerations to a pre-relativistic treatment, we attempt to compute the electromagnetic mass obtaining the equation of motion of an 'electron-like' extended charged particle. Finally on the basis of the convective derivative terms we attempt to elucidate their physical meaning. It is worthy to recall here that a long-time ago Heaviside (1893) emphatically expressed the idea that "It seems [...] not unlikely that in discussing purely electromagnetic speculations, one may be within a stone's throw of the explanation of gravitation all the time".

## THE ELECTROMAGNETIC ORIGIN OF INERTIA PROPERTY

The inertia force has remained a mystery ever since it was described by Newton, and up until now there has been no straightforward clear explanation for it. Newton's first law, the law of inertia, states that a body remains at rest or in motion with the same speed and in the same direction unless acted upon by a force. From Newton's second law of motion we know that, to overcome inertia, the applied force has to have the magnitude of the inertia force. So, despite knowing that for every action (acceleration) there is a reaction (inertia) as stated by the Newton's third law, these two forces do not cancel each other since velocity has to change for the effect to take place due to the retarded fields emanating from an accelerated charge.

In the Lagrangian formalism of a charged particle the generalized (canonical) momentum must be $\mathbf{p} = m\mathbf{v} + q\mathbf{A}$. Whenever the particle is not subject to an external force, it is $\mathbf{p}$ rather than $m\mathbf{v}$ that is conserved. Maxwell advocated in 1865 that the vector potential could be seen as a stored momentum per unit charge, and Thomson in 1904 interpreted $\mathbf{A}$ as a field momentum per unit charge. More recently, Mead (1997) derived standard results of electromagnetic theory of the direct interaction of macroscopic quantum systems assuming solely the Einstein-de Broglie relations, the discrete nature of charge, the Green's function for the vector potential, and the continuity of the wave function - without any reference to Maxwell's equations. Holding an opposite view are Heaviside and Hertz who envisaged in the vector potential merely as an auxiliary artifact to computation (Semon and Taylor, 1996; Coïsson, 1973; Konopinski, 1978; Calkin, 1979; Gingras, 1980; Jackson and Okun, 2001; Iencinella and Matteucci, 2004; Fowles, 1980).

In this paper inertia is discussed in terms of the "potential momentum", or vector potential created by the particle, as the primary source for the inertia force. It is known that any charged particle in motion constitute an electric current with an associated "potential momentum", $\mathbf{A}$. When the velocity is uniform, $\mathbf{A}$ is constant in magnitude and no "potential momentum" will be exchanged between the field and the particle. If the velocity varies, however, the difference in "potential momentum" caused by the resulting acceleration will exert a force on the particle itself which will be opposed to the external applied force.

Fig. 1 shows an accelerated positive charge with the respective star-like electric field lines $\mathbf{E}_+$ represented when at rest; to the right we have pointed the acceleration vector, $\mathbf{a}$, and velocity vector, $\mathbf{v}$; above we have the current density, $\mathbf{J}$, and the vector potential, $\mathbf{A}$, directed along the particle's movement. Also, it is represented the induced electric field, $\mathbf{E}_i$, as given by the equation $E_i = -\partial A/\partial t$. Jefimenko (2000) refers to $\mathbf{E}_i$ as the electrokinetic field. At the bottom it is the induced electric force $F_{E_i} = qE_i$ that acts on the charge and also the inertia force, $F_I$, that is, the force produced by the reaction of a body to an accelerating force. The attraction and repulsion signs refer to the forces that the particle "feels" due to the interaction between its own electric field and the induced electric field $\mathbf{E}_i$. This electric field acts to counteract the acceleration of the charge, and exists only during the acceleration time of the particle. We clearly see that $F_{E_i}$ has the same direction as the inertia force $F_I$, but does it has the same magnitude? Examining Fig. 1 one can readily see why a particle "feels" an inertial force whenever it is submitted to accelerating or decelerating external forces. When in motion it generates a current $I$ (and a related vector density of charge $\mathbf{J}=\rho\mathbf{v}$) and a potential vector $\mathbf{A}$ in the same direction of velocity, the retarded field is given by:

$$A(x,t) = \frac{\mu_0}{4\pi} \iiint_V \frac{[J(x',t')]_{ret}}{r} dx', \qquad (1)$$

with $r = |x - x'|$ and $t' = t - r/c$. As the current must be evaluated at the retarded time we follow a formalism developed by Lorentz to understand the action of each part of a particle on the others since we assume it is not punctual (see Jackson and Okun (2001) for more details on the self-reaction force). The retarded quantity has an expansion in Taylor's series:

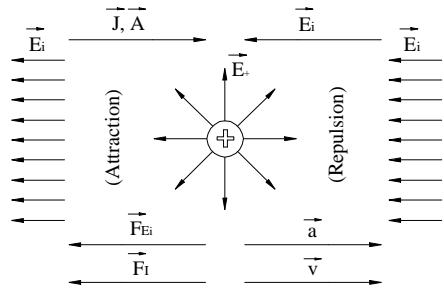

**FIGURE 1.** Schematic of an Accelerated Positive Charge Illustrating the Origin of the Inertia Force.

$$[\ldots]_{ret} = \sum_{n=0}^{\infty} \frac{(-1)^n}{n!} \left(\frac{r}{c}\right)^n \frac{\partial^n}{\partial t^n} [\ldots]_{t=t'}. \qquad (2)$$

This is still a pre-relativistic formulation restraining the validity of its results to low particle velocities. Now, we decompose the total fields into the external field $\mathbf{A}_{ext}$ and the self-fields $\mathbf{A}_s$:

$$\mathbf{A} = \mathbf{A}_{ext} + \mathbf{A}_s. \qquad (3)$$

The total linear momentum is conserved only when using the canonical momentum (Landau and Lifschitz, 1989), $\mathbf{p}$, and it is given by:

$$\mathbf{p} = \mu\mathbf{v} + q\mathbf{A}. \qquad (4)$$

A possible approach to the problem consists in using Newton's second law in the non-relativistic limit for a charge $q$ in the presence of an external force $\mathbf{F}^{ext}$. This is a continuity fluid-like equation:

$$\frac{d\mathbf{p}}{dt} = \overline{m}\frac{d\mathbf{v}}{dt} = \mu\frac{d\mathbf{v}}{dt} + q\frac{D\mathbf{A}}{Dt} = \mathbf{F}^{ext}, \qquad (5)$$

where the observable mass is $\overline{m}$ and the mechanical (bare) rest mass is $\mu$. Substituting the particle acceleration $\mathbf{a} = d\mathbf{v}/dt$ into Eq. 5, it leads to:

$$\mu \boldsymbol{a} = \boldsymbol{F}^{ext} - q\frac{D\boldsymbol{A}}{Dt}. \tag{6}$$

Here, *D/Dt* means the total (convective) derivative. Its use seems to offer a natural frame to describe the motion of an electromagnetic system relatively to an inertial frame. Maxwell expressed the electromotive force (Jackson and Okun, 2001) as **E** = -*D***A**/*Dt* although he did not explore fully its consequences more carefully studied by others after him (Searle, 1896; Hertz, 1962; Rosen and Schieber, 1982; Pinheiro, 2006). The convective derivative operator in space is given by:

$$\frac{D}{Dt} = \frac{\partial}{\partial t} + \boldsymbol{v} \cdot \frac{\partial}{\partial \boldsymbol{r}}. \tag{7}$$

After its substitution into Eq. 6, we obtain:

$$\mu \boldsymbol{a} = \boldsymbol{F}^{ext} - q\frac{\partial \boldsymbol{A}}{\partial t} - q\boldsymbol{v} \cdot \frac{\partial \boldsymbol{A}}{\partial \boldsymbol{r}}. \tag{8}$$

This result is important because it tells us that when a particle is acted upon by an external force and accelerates, the change in "potential momentum", **A**, acts oppositely in order to conserve linear momentum. The magnitude of the force derived by this change - the *q D***A**/*Dt* term - maybe interpreted as an induced force of inertia acting on the particle. Since $\boldsymbol{E}_i = -\partial \boldsymbol{A}/\partial t$ when taking due care with of the convective derivative, remark that we can rewrite Eq. 8 also in the form:

$$\mu \boldsymbol{a} = \boldsymbol{F}^{ext} - q\frac{\partial \boldsymbol{A}}{\partial t} + q[\boldsymbol{v} \times \boldsymbol{B}] - q\nabla_r(\boldsymbol{v} \cdot \boldsymbol{A}), \tag{9}$$

where the B-field appears explicitly. The terms with the self-fields give the reaction force and as well terms of higher order with no clear physical interpretation (see Jackson and Okun (2001), sec.17.3). The last term in Eq. 9 is related to Aharonov-Bohm effect (Boyer, 1973; Aharonov and Bohm, 1959; Trammel, 1964). We can see that the particle's own Coulomb field doesn't contribute to a net self-force; when subject alone to its own Coulomb field the extended particle describes a uniform velocity motion. Now we can apply Eq. 8 to an extended charged particle while assuming spherical distribution of charge and slow acceleration. These assumptions probably describe well a charged particle at small velocities. At higher velocities the particle acquires an ellipsoidal shape and our approximation are not anymore valid. Using Lorentz's procedure we can conveniently write Eq. 8 under the form:

$$\mu \boldsymbol{a} = \boldsymbol{F}^{ext} - \int d^3x \rho(x,t)\frac{\partial \boldsymbol{A}_s(x,t)}{\partial t} - \int d^3x \rho(x,t)\left(\boldsymbol{v}(t) \cdot \frac{\partial \boldsymbol{A}_s(x,t)}{\partial \boldsymbol{r}}\right). \tag{10}$$

Now we can search for terms with interesting physical meaning by inserting Eq. 2 into Eq. 10. The first integral in the right hand side gives the following serial development:

$$\boldsymbol{I}_I^n = -\frac{1}{4\pi\varepsilon_0}\sum_{n=0}^{\infty}\frac{(-1)^n}{n!\,2c^{n+2}}\frac{\partial^{n+1}\boldsymbol{v}}{\partial t^{n+1}}\int d^3x \int d^3x' \rho(x,t) r^{n-1} \rho(x',t). \tag{11}$$

The first two terms of the series are, respectively,

$$\boldsymbol{I}_I^0 = -\frac{U_{es}}{c^2}\boldsymbol{a}, \tag{12}$$

and:

$$I_1^1 = \frac{e^2}{2c^3}\frac{\partial^2 \mathbf{v}}{\partial t^2} = \frac{e^2}{2c^3}\dot{\mathbf{a}}. \tag{13}$$

Together, they constitute the radiation reaction field. Millonni (1984) has shown that from the fluctuation-dissipation theorem it must exists an intimate connection between radiation reaction and the zero-point field (ZPF), since the spectrum of the ZPF depends of the third derivative of the particle's position vector. Here, $e^2 = q^2/(4\pi\varepsilon_0)$, and $R$ is the classical particle radius (Leighton and Sands, 1964). It is important to point out that a factor 1/2 has to be inserted above into the integrals appearing in Eq. 10 since they represent the interaction of a given element of charge $dq$ with all the other parts, otherwise we count twice that reciprocal action. Recall that the value of the electrostatic energy is given by:

$$U_{es} = \frac{1}{2}\int d^3x \rho(x,t)\Phi(x,t), \tag{14}$$

Where:

$$\Phi(x,t) = \int d^3x' \frac{\rho(x',t)}{4\pi\varepsilon_0 r}, \tag{15}$$

represents the instantaneous electrostatic potential. The obtained value is related to the assumed structure of the "electron-like" particle with the charge concentrated on the surface of a sphere with radius $R$ (Weisskopf, 1949; Rohrlich, 1960), while if we assume a charged spherical particle we should obtain instead $U_{es} = 2e^2/3c^2R$. The electron is likely to be hollow since otherwise a singularity exists at the center of the sphere which amount to an infinite energy inside; there is no electric field inside the classical radius. See *a propos* the clear discussion about this matter in Jackson and Okun (2001), Marmet (2003) and Dirac's "bubble-model" of the electron (Dirac, 1962). Finally, let's consider the second integral in the right-hand side of Eq. 8 given by:

$$I_2 = -\int d^3x \rho(x,t)\left(\mathbf{v}\cdot\frac{\partial \mathbf{A}_s}{\partial \mathbf{r}}\right). \tag{16}$$

Applying again the Lorentz's procedure we then have:

$$I_2^n = -\frac{1}{2}\sum_{n=0}^{\infty}\frac{(-1)^n}{n!\,c^{n+2}}\int d^3x\rho(x,t)\int d^3x'\frac{r^{n-2}}{4\pi\varepsilon_0}\frac{\partial^n \rho(x',t)\mathbf{v}(t)}{\partial t^n}(\mathbf{v}(t)\cdot\mathbf{u}_r), \tag{17}$$

with $\mathbf{u}_r$ denoting the unitary radius vector. The first two terms of the previous power expansion are:

$$I_2^0 = \frac{v^2}{c^2}\mathbf{F}_{es}, \tag{18}$$

which gives a null result for a spherical symmetry, and:

$$I_2^1 = -\frac{1}{2c^3}\int d^3x\rho(x,t)\frac{\partial}{\partial t}\left[\mathbf{v}(t)\int d^3x'\frac{\rho(x',t)}{4\pi\varepsilon_0 r^2}(\mathbf{v}(t)\cdot\mathbf{u}_r)\right]. \tag{19}$$

The $n = 3$ term is of order of the second derivative $\sim \partial^2 v/\partial t^2$, negligible when compared to the previous ones under our initial assumptions. We postpone the discussion of the magnetic component of the self-force to Sect. 3.

So far we can state that whenever there is a particle with mass *m* and charge *q* accelerating or decelerating it will be generated an opposed force given by $F_{E_i}$ which will act against the acceleration vector. This mechanism derives from the exchange of "potential momentum" between the particle and the field generated by its motion. We recover the total particle mass as the sum of the mechanical mass term (which we assume as hypothetically generated by interactions of other nature than electromagnetic) plus the mass of electromagnetic origin (which results from the time-dependent **A** vector). Finally, the Newton equation of motion of an accelerated "electron-like" extended charge is therefore as a first approximation given by (a dot means a time derivative)

$$\dot{v}\overline{m} = F^{ext} + F_{Sch} + F_{st}. \quad (20)$$

Here,

$$\overline{m} = \mu + \frac{U_{es}}{c^2} = \mu + \frac{e^2}{2Rc^2}, \quad (21)$$

is the observed rest mass of the charged sphere, while the Schott term (source of the radiation field) is given by (Schott, 1912):

$$F_{Sch} = \frac{e^2}{2c^3}\dot{a}, \quad (22)$$

and the last term:

$$F_{st} = -I_2^1, \quad (23)$$

is the stress force (source of the deformation of the field lines). It is worth to point out that in Eq. 20 Lorentz covariance is recovered. This question has been analyzed by Hnizdo, in particular showing the contributions of the electromagnetic self-field to the energy and momentum of a charge and/or current carrying body and the important role of hidden mechanical momentum (Hnizdo, 1997). Moreover, it results from the previous developments that the electromagnetic mass $m_{em}$ equals the electrostatic mass $m_{es} = e^2/2Rc^2$ instead of equaling 4/3 the electrostatic mass, as in the Lorentz-Abraham force and power equations (Arzeliés, 1966; Yaghjian, 1992; Poincaré, 1989). This result supports the claim that all the electron mass as well as its inertia has an entirely electromagnetic origin, as it is well-known, a viewpoint defended by Lorentz (1992) and Schott (1912). The investigations of Fermi (1923) using a variational method also have shown the entire electromagnetic origin of the electron inertial mass. The electromagnetic interaction energy can explain why under the effect of an external force acting on the charged particle behave as if it possessed a mass $m_{es}$. In fact, the discrepancies found in literature are due to faulty electromagnetic momentum- and energy-density expressions (Möller, 1949; Butler, 1969; Wilson, 1936).

Within this framework there is no contribution of other kind of interactions to inertia except purely electromagnetic interactions. In accordance, the mechanical mass must be null, $\mu = 0$. This result brings some convenience since whenever we calculate the electron mass with the classical electron radius (which is, however, an adjustable value dependent on the model, see, for example, Becker (1964), the expression obtained for the electrostatic mass gives exactly the experimental value of the electron mass at rest.

Now returning to Eq. 10 it is clear that the inertial force is composed basically of two components: i) the local time derivative of the vector potential and, ii) the convective term on **A**. The term related to a local change (the time derivative) of **A**, as we will see, represents a resistance to change induced by the charge acceleration due to the action of its immersed own field (the reader is referred to Campos and Jimenez (1989) for a similar interpretation). In fact, the effect of the self-field on the charged particle can be well understood. When an electron suddenly decelerates, the magnetic field increases. According to the induction law, however, an increasing magnetic field gives rise to an electric field. And this same electric field will act on the electron, accelerating it. This effect is interpreted as a contribution to inertia. Moreover, the serial developments effectuated above suggest that the electrokinetic force is by itself the source of the inertial mass and of the radiation reaction force (or *Schott term*). The radiation reaction force contributes to inertia through transfers of energy back and forth between the field and the source (due to the action of the source at the retarded time on itself, see also Yaghjian (1992)). But the electrokinetic force term, which represents a *local* time derivative of $\mathbf{A_s}$, means that the mass is a locally determined quantity, weakening Mach's conjecture. As long as the electrokinetic term is the mass generator we can regard $\mathbf{E_i}$ as the source of the kinetic energy of the particle. It is possible that this local reaction force acting on the charge could be in fact, according to Newton's third law, a reaction from the *physical vacuum*, since the experimental findings by Graham and Lahoz (1980) implies that the vacuum is the seat of "something in motion", like Maxwell himself envisaged the "aether". And it was shown that a medium in uniform motion with velocity $\mathbf{v}$ plays the role of the vector potential while the charge is proportional to Fresnel's dragging coefficient for light in moving media (Leonhardt and Piwnicki, 1999).

The last term of Eq. 10 represents the convective term on $\mathbf{A}$ and it is non-null when the vector potential, the field source, is inhomogeneous. It measures the acceleration of the extended charged particle relatively to its own vector potential field lines located ahead of the charged particle line of motion. The physicality of the vector potential is now well proven experimentally (Tonomura, 2005). Dirac understood that there are nowadays good reasons to postulate the existence of an aether needing an "elaborate mathematics for its description" (Dirac, 1962) and he attempted to build a new classical theory of electrons not based on gauge transformations. According to Dirac (1951) the velocity $v$ appearing in the vector potential has the physical significance of the velocity with which an electron charge must flow in the ether. It is remarkable how progressively mathematical sophistication is being introduced to build a differential structure of space-time, beginning with Maxwell when he identified the vector potential with Faraday's intuitive idea of an electro-tonic state which now developed into a precise mathematical concept of the connection on a fiber bundle.

For the purpose to clarify further the role of the stress force let's consider a moving electron along the x-axis. The component of this force term is opposed to the direction of the acceleration acting effectively as a *radiation reaction force* (Rohrlich, 2000). In this case it is easily obtained the component of this stress force:

$$I_{2x}^1 = -\frac{e^2}{2Rc^2}\frac{v_x}{c}\frac{\partial v_x}{\partial t}. \tag{24}$$

Now we can calculate the power consumption multiplying this (stress) force by the velocity $v$ at time $t = \Delta t$, such as $v = a\Delta t$, and also taking into account that during this interval of time the particle was displaced by $R = v\Delta t$ in this space (medium) where the field lines are build up. We thus obtain the power radiated by an accelerated charge (Larmor's formula):

$$P_s = -F_s v = \frac{e^2 a^2}{2c^3}. \tag{25}$$

This finding is consistent with the physical reinterpretation advanced by Harpaz and Soker (1998; 2003) which explains the emitted radiation by an accelerated charge as due to the relative acceleration between the electric charge and its own electric field lines that do not move with the charge, instead of the usual argument of the emitted radiation as due to a relative acceleration between the charge and an observer. Therefore, it is this stress force that is the cause for resistance to acceleration. But, contrary to Harpaz and Soker approach, we obtain this drag force in a consistent manner through the continuity equation for canonical momentum applied to an extended charge. Our result is consistent with the mechanism of change in the inertial mass of a system of point charges interpreted by

Boyer (1973; 1978; 1979) as being due to the electromagnetic fields change during the acceleration, in such a way that each charge causes a new electromagnetic force on the other.

The convective derivative introduced in the continuity equation for momentum flux (see Eq. 10) traduces not only the conversion from potential to kinetic energy (term $\mathbf{I_1}$), but also the convection of potential electromagnetic momentum (term $\mathbf{I_2}$), the true flux of electromagnetic momentum through the medium, which is in fact related to a deformation of the vector potential in space. So to use an analogy between optics in fluids and the gravitational field (Leonhardt and Piwnicki, 1999), we sustain that the charged particle inside the flux of the vector potential acts as submitted to inertial forces.

## INERTIA MANIPULATION

In the previous developments we supposed the particle velocity small and assumed spherical symmetry. But considering more complex particle structures could lead us to other substantial findings. For example, Bohm and Weinstein (1948) hypothesized the oscillations of the particle's own charge distribution and showed that when these oscillation were quantized the energy of the first excited state is of the order of the meson self-energy. Pekeris (1975) built a hydrodynamical model of the electron where instead of Poincaré stresses (Poincaré, 1989) the repulsive forces between charges were held by a dynamic pressure generated by a steady circulation of a nonconducting perfect fluid. Therefore, the concept of electromagnetic mass has far-reaching aspiration laying the foundations for the theory of elementary particles.

Indeed, other proposals to control inertia through appropriate external fields have been advanced, although on different grounds than ours. For instance, Woodward (1990; 1991) demonstrated theoretically that by energy flowing to a charged capacitor, Sciama's (1953; 1964) inertia term acquires quite large transient contributions, proportional to $1/G$ ($G$ denotes the gravitational constant). Also, it can be shown that the equations of classical electrodynamics for an electric dipole through the self-retarded field interactions allow a self-sustained motion, even in the absence of external fields (Cornish, 1986; Griffiths, 1986; Petkov, 1998). Moreover, it was shown that a system consisting of two point charges, under the action of a weak gravitational field has an increase of weight (Boyer, 1973; 1978).

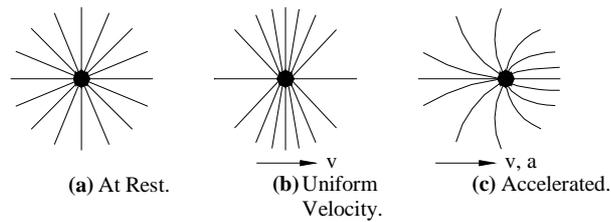

(a) At Rest.  (b) Uniform Velocity.  (c) Accelerated.

**FIGURE 2.** Electric Field of a Charged Particle.

As the foregoing discussion of the last Section was limited to small velocities ($v << c$), it did not contemplate the magnetic field term in Eq. 9 which allows an interesting physical interpretation, as we will see now. According to the arguments presented by Marmet (2003) and anticipated by Heaviside (1889) the Biot-Savart law is valid for a charge current but when considering an individual charged particle moving with velocity $v$ its self-magnetic field can be supposed to be isotropically distributed around the particle, as it happens to the electric field around the particle at rest. Along this line of reasoning we assume that the magnetic field has an absolute value given by:

$$B_s = \frac{\mu_0}{4\pi} \frac{qv}{r^2}. \qquad (26)$$

The integration in volume gives the associated magnetic energy:

$$W_{mag} = \frac{1}{2\mu_0}\int d^3x B_S^2 = U_{es}\frac{v^2}{c^2}. \qquad (27)$$

The above equation shows that the mass increase due to the self-magnetic field is of second order in $v^2/c^2$ and supports the view that the relativistic increase of mass with velocity is due to the associated magnetic field resulting from the action of the particle's self-field (Marmet, 2003; Breitenberger, 1968), although its effects are only noticeable at high velocities.

The difference between these two referred reaction forces (the second and third terms on the right-hand side of Eq. 10) can be better understood through the electric field lines of a charge when at rest, with a uniform velocity or with acceleration, as illustrated in Fig. 2 and drawn according to the equations obtained by Singal (1997). At rest the electric field lines are spherically symmetric around the center of the particle. With a constant velocity, the electric field lines tend to the plane perpendicular to the velocity vector, in a symmetric fashion, while when in acceleration the electric field lines deform to the opposite direction of the acceleration vector, as shown in Fig. 2.c). From Eq. 26 we obtain:

$$vB_s = E\frac{v^2}{c^2}, \qquad (28)$$

since $E = q/4\pi\varepsilon_0 r^2$ for a particle at rest. After multiplying Eq. 18 by the charge density and next integrating in space we thus find Eq. 28. Notice that $vB_S$ represents an induced electric field by the particle on itself augmenting the electric field perpendicular to the velocity vector. This effect can explain the "compression" of the electric field lines with velocity and as well the relativistic increase of mass. This makes us see a "polarization" effect on the charge and its internal structure, since with increased velocity the electric field lines tend to be oriented progressively into a flat disc perpendicular to the particle trajectory. Eq. 28 shows that the induced electric field depends on velocity, being equal in magnitude to the rest electric field of the particle whenever $v = c$. Thus Eq. 18 is not null when the velocity is different from zero, as can be seen by the strong field asymmetry shown in Fig. 2.b) which happens at relativistic velocities.

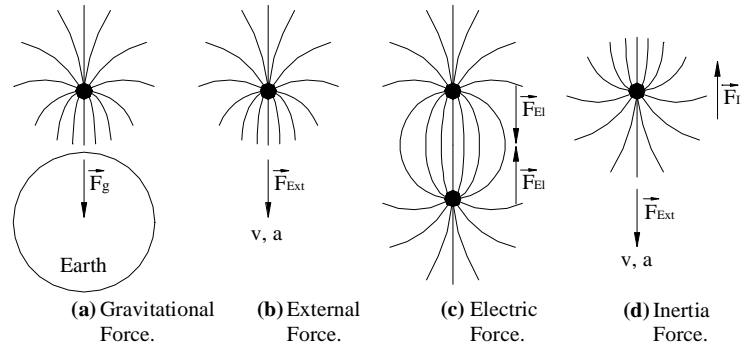

(a) Gravitational Force.    (b) External Force.    (c) Electric Force.    (d) Inertia Force.

**FIGURE 3.** Electric Field Deformation of a Charged Particle Subject to Different Forces.

It is possible that the compression of the electric field lines by this effect of internal polarization may possibly reduce the electron radius by decreasing the volume occupied by its electrostatic energy, and thereby increasing its mass (Miller, 2003). This could explain the relativistic increase of mass and as well as the impossibility of surpassing the light velocity in these conditions, since ultimately a black hole would be created (Burinski, 2005).

For an accelerating charge, we have seen that it is generated an induced electric field given by the time variation of the potential vector. Since inertia is given by the forces produced on the particle by the vector potential self-induced field, these can be nulled out if a charge is manipulated with an unbalanced convective derivative of the potential vector. It will not feel an inertia reaction force but only a directed acceleration field. Our treatment of the problem makes it possible to find a way to control inertia by an appropriate interplay between the external and self-vector potential. Manipulating the particle directly with an external change of the potential vector that exactly cancels or surpasses the convective derivative of its own potential vector, then the particle will be manipulated without inertia. For example, Pierce (1955) has shown that when a charge moves along a nondispersive transmission line its electromagnetic mass at velocities greater than the wave velocity is negative. Also, when studying the system of two charged point particles traveling side by side, Boyer (1978; 1979) proved that during acceleration the electromagnetic fields change in order that each particle will act with a new electromagnetic force on the other, exactly balancing the external force accelerating the system and which represent a change in the inertial mass of the system. Proposals to manipulate the electromagnetic quantum vacuum and/or its interaction with matter in order to modify inertial and gravitational forces have been advanced (Haisch, Rueda and Puthoff, 1994; Haisch and Rueda, 2000; 2001; Puthoff, 1989; 2002).

Fig. 3 illustrates the action of external forces of several kinds producing an asymmetry in their electric field distribution in order that the sink of the flux lines are oriented to the acceleration vector. We wish to point out that, although the deformation of the electric field lines due to the inertia force are represented in Fig. 3.d), the effective/observable lines of force are as shown in Fig. 3.b) since the external force applied is superior to the inertia force and provides the work to deflect the lines of force.

In neutral particles the electric fields of each charge continue to exist and thus each particle will be subject to inertia forces by the self induced local electric field appearing due to the acceleration, although the net macroscopic induced electric field will be zero. The magnetic fields produced by neutral matter in motion cancel out macroscopically but always contain all the magnetic energy generated (Marmet, 2003). Therefore, the relativistic mass increase is not easily prone to control, since any external magnetic field that we could apply on a particle to cancel its own self generated magnetic field would only sum-up.

Probably the best way to control inertia is *engineering the vacuum* (Lee, 1988), through the manipulation of the basic constants of nature: the magnetic permeability $\mu_0$ and electric permittivity $\varepsilon_0$. In order for the particle to generate a zero magnetic field it is sufficient to make $\mu_0 = 0$. As we have shown, the mass of the particle depends on the electrostatic energy of its field and on the velocity of light in the surrounding space (through the mass-energy relationship). If we succeed by operational means to make $c \rightarrow \infty$ we would have zero mass for the particle and thus no inertia. The vacuum modification effect on the speed of light by the Casimir effect, now known as the Scharnhorst effect (Davies, 2004) is a serious candidate to accomplish this goal by means of a modified vacuum energy density. The manipulation of vacuum is of vital importance both in surpassing inertia and producing propulsion, since space gradients of different refractive indexes act as attractive or repulsive gravitational fields (Puthoff, 2002). Further research is necessary to know exactly how to manipulate $\varepsilon_0$ and $\mu_0$ in order to change the local velocity of light and vacuum energy density. The theory developed here, however, provides a framework on how this manipulation can affect inertia since a link was shown to electric and magnetic fields and these depend on the local speed of light.

## CONCLUSION

In the frame of a simple and intuitive approach, using Newton's second law, we obtain in a covariant form the dynamical equation of motion of an extended charged particle, subject to the Lorentz's procedure with retarded fields. It is shown that the electron mass has a complete electromagnetic origin. This explanation of inertia in terms of the vector potential turns out to give a more appropriate framework to understand phenomena related to the classical electromagnetic mass theory. Our research provides significant insight over the origin of inertia as due to the time and space changes (acceleration and velocity dependent term, respectively) of the particles own vector potential and we link these components with the Schott term and a stress force exerted by the charge on the lines of force. We offer possible means of inertia control, with important applications for electromagnetic and future space propulsion. In addition, we believe this work contributes one step further in the desired unification of electromagnetic and gravitational forces.

## NOMENCLATURE

| | | | | |
|---|---|---|---|---|
| $q$ | = particle's charge (C) | | $\mathbf{A}$ | = vector potential (T.m) |
| $\overline{m}$ | = observable particle's mass (kg) | | $\mathbf{E}_i$ | = induced electric field (or electrokinetic field) (V/m) |
| $m_{es}$ | = electrostatic mass (kg) | | $\mathbf{F}_{Ei}$ | = induced electric force (N) |
| $m_{em}$ | = electromagnetic mass (kg) | | $\mathbf{F}_I$ | = inertia force (N) |
| $R$ | = classical particle's radius (m) | | $I$ | = electric current (A) |
| $\mu$ | = mechanical (bare) particle's rest mass (kg) | | $\mathbf{J}$ | = vector density of charge (A/m$^2$) |
| $\mu_0$ | = permeability of free space (H/m) | | $\mathbf{B}$ | = magnetic field (T) |
| $\varepsilon_0$ | = permittivity of free space (F/m) | | $U_{es}$ | = particle's electrostatic energy (J) |
| $c$ | = speed of light (m/s) | | $\Phi(\mathbf{x},t)$ | = instantaneous electrostatic potential (V) |
| $\mathbf{a}$ | = acceleration (m.s$^2$) | | $\mathbf{F}^{ext}$ | = external force (N) |
| $\rho$ | = charge density (C/m$^3$) | | $\mathbf{F}_{Sch}$ | = force Schott term (N) |
| $\mathbf{v}$ | = velocity of the particle (m/s) | | $\mathbf{F}_{st}$ | = stress force (N) |
| $\mathbf{p}$ | = generalized (canonical) momentum (kg.m/s) | | $W_{mag}$ | = magnetic energy (J) |

## ACKNOWLEDGMENTS

We would like to thank partial support from the Rectorate of the Technical University of Lisbon and Fundação Calouste Gulbenkian.